\documentclass[preprint,12pt,authoryear]{elsarticle}
\usepackage{graphicx}
\usepackage{amsmath,amssymb,amsfonts,longtable,rotating}

\usepackage{natbib}
 \usepackage[margin=1in]{geometry}

\newtheorem{example}{Example}[section]

\makeatletter
\def\ps@pprintTitle{%
 \let\@oddhead\@empty
 \let\@evenhead\@empty
 \def\@oddfoot{\centerline{\thepage}}%
 \let\@evenfoot\@oddfoot}
\makeatother

\usepackage{tikz}
\usetikzlibrary{plotmarks}

\usepackage{verbatim}

\journal{Communications in Statistics - Simulation and Computation }
\begin{document}

\begin{frontmatter}

\title{The ordering of future observations from multiple groups}

\author{Tahani Coolen-Maturi\corref{cor}}
\cortext[cor]{Corresponding author}
\ead{tahani.maturi@durham.ac.uk}

\address{Department of Mathematical Sciences, Durham University, Durham, DH1 3LE, UK}

\begin{abstract}
There are many situations where  comparison of different groups is of great interest. Considering the ordering of the efficiency of some treatments is an example. We present nonparametric predictive inference (NPI) for the ordering of real-valued future observations from multiple independent groups. The uncertainty is quantified using NPI lower and upper probabilities for the event that the next future observations from these groups are ordered in a specific way. Several applications of these NPI lower and upper probabilities are explored, including multiple groups inference,  diagnostic accuracy and ranked set sampling. 
\end{abstract}

\end{frontmatter}


\section{Introduction}\label{intro}
There are many situations where  comparison of different groups is of great interest. For example, comparing the effectiveness of different treatments, and whether they are ordered in a specific way. In classical tests, one may want to test the null hypothesis  that the location parameters  of different populations  are equal, against different alternatives, e.g.\ 
Kruskal-Wallis  test of this null hypothesis against the alternative that at least one of them is not equal, and the Jonckheere-Terpstra  test of this null  hypothesis against a specific ordered alternative. Several nonparametric tests for the  ordered alternative problem are introduced in the literature, we refer the reader to \citet{TM03} for an overview of these tests. Later in the paper,  we will compare the proposed method with some of these nonparametric tests. Another interesting application is  ranked set sampling  \citep{McIn52}, which is often considered as an alternative to the simple random sampling when the measurement of the characteristic of interest is costly and time consuming \citep{LB08}. The  ranked set sampling's  inferences are often based on the assumption of  perfect ranking of the samples. Finally, the volume under  the  receiver operating characteristic surface is commonly used as an overall measure of the accuracy of diagnostic tests \citep{NC14}. The volume is defined as the probability that  the test results for different groups are perfectly ordered. In classical statistics, the focus is mainly on estimation and hypothesis testing, while in many applications it maybe attractive to quantify the uncertainty about  future observations.

In this paper, we present nonparametric predictive inference (NPI) for the ordering of real-valued future observations from multiple independent groups. The uncertainty is quantified using NPI lower and upper probabilities for the event that the next future observations from these groups are ordered in a specific way. Several applications of the NPI lower and upper probabilities are explored, including multiple groups inference,  diagnostic accuracy and ranked set sampling. A brief overview of NPI is give below.

Nonparametric predictive inference (NPI) \citep{AC04,CF06} is based on the assumption $A_{(n)}$ proposed by  \citet{HB68}. Let $X_1,\ldots,X_{n},X_{n+1}$ be real-valued absolutely continuous and exchangeable random quantities. Let the ordered observed values of $X_1,X_2,\ldots,X_{n}$ be denoted by $x_{1}<x_{2}<\ldots<x_{n}$ and let $x_{0}=-\infty$ and $x_{n+1}=\infty$ for ease of notation. We assume that no ties occur; ties can be dealt with in NPI as in \citet{CF06}. For $X_{n+1}$, representing a future observation, $A_{(n)}$ partially specifies a probability distribution by 
$P(X_{n+1} \in (x_{i-1},x_{i})) = \frac{1}{n+1}$ for $i=1,\ldots,n+1$. Inferences based on $A_{(n)}$ are predictive and nonparametric, and can be considered suitable if there is hardly any knowledge about the random quantity of interest, other than the $n$ observations, or if one does not want to use such information. $A_{(n)}$ is not sufficient to derive precise probabilities for many events of interest, but it provides bounds for probabilities via the `fundamental theorem of probability' \citep{DF74}, which are lower and upper probabilities in imprecise probability theory \citep{TAFC14}. 
\citet{AC04} proved that NPI has strong consistency properties in the theory of imprecise probability \citep{TAFC14}, it is also exactly calibrated from frequentist statistics perspective \citep{LF05}. In NPI, uncertainty about the future observation $X_{n+1}$ is quantified by lower and upper probabilities for events of interest.The NPI lower and upper probabilities are the sharpest bounds on a probability for an event of interest when only $A_{(n)}$ is assumed. Informally, $\underline{P}(A)$ ($\overline{P}(A)$) can be considered to reflect the evidence in favour of (against) event $A$.

While it is natural to consider inference for a single future observation in many situations, one may also be interested in multiple future observations. This is possible in NPI in a sequential way, taking the inter-dependence of the multiple future observations into account \citep{ACL04}. In this paper, attention is restricted to a single future observation per group, leaving generalisation to multiple future observations as an interesting challenge for future research.  NPI has been introduced for pairwise and multiple comparisons for different data types; including  binary, ordinal data, real-valued and right censored data, see e.g.\ \cite{CFPCS07,CCSCME13,CF96,CoCC2012a}. However, those approaches did not consider the ordering of these groups. In this paper, we present NPI lower and upper probabilities for the event that the next future observations from these groups are ordered in a specific way. Through this paper we assume that the  groups are fully independent, in the sense that any information about one group, does not provide any information about the other group.

The  paper is organised as follows.  Section \ref{sec.main} introduces  the main results of NPI for the ordering of future observations from multiple groups.  Section \ref{sec.app} presents three applications of the proposed method including multiple groups inference,  diagnostic accuracy and ranked set sampling. Examples are provided throughout for illustration purposes. The paper ends with concluding remarks in Section \ref{sec.con}, which includes some related future research challenges.


\section{The ordering of future observations from multiple groups}\label{sec.main}

\subsection{The ordering of three future observations}
In this section we  consider three independent groups, $X$, $Y$ and $Z$, and the aim is to introduce  the NPI lower and upper probabilities for the event $X_{n_x+1}<Y_{n_y+1}<Z_{n_z+1}$, where $X_{n_x+1}$, $Y_{n_y+1}$ and $Z_{n_z+1}$ are the next future observations from these groups, respectively.   In order to introduce NPI for such an event we will apply $A_{(n)}$ per group, so we need to introduce the following notations.

From group $X$ we have ordered observations $x_{1}<x_{2}<\ldots<x_{n_x}$ and let $x_{0}=-\infty$ and $x_{n_x+1}=\infty$ for ease of notation.  For $X_{n_x+1}$, representing a future observation from group $X$, $A_{(n_x)}$ partially specifies a probability distribution by 
$P(X_{n_x+1} \in (x_{i-1},x_{i})) = \frac{1}{n_x+1}$ for $i=1,\ldots,n_x+1$. Similarly, let the ordered observations from group $Y$  be denoted by $y_{1}<y_{2}<\ldots<y_{n_y}$ and let $y_{0}=-\infty$ and $y_{n_y+1}=\infty$ for ease of notation.  For $Y_{n_y+1}$, representing a future observation from group $Y$, $A_{(n_y)}$ partially specifies a probability distribution by $P(Y_{n_y+1} \in (y_{j-1},y_{j})) = \frac{1}{n_y+1}$ for $j=1,\ldots,n_y+1$. And finally, let the ordered observations from group $Z$ be denoted by $z_{1}<z_{2}<\ldots<z_{n_z}$ and let $z_{0}=-\infty$ and $z_{n_z+1}=\infty$ for ease of notation.  For $Z_{n_z+1}$, representing a future observation from group $Z$, $A_{(n_z)}$ partially specifies a probability distribution by 
$P(Z_{n_z+1} \in (z_{k-1},z_{k})) = \frac{1}{n_z+1}$ for $k=1,\ldots,n_z+1$.\\

The NPI lower and upper probabilities for the event $X_{n_x+1}<Y_{n_y+1}<Z_{n_z+1}$ are 
\begin{align}
\hspace{-3mm}\underline{P}(X_{n_x+1}<Y_{n_y+1}<Z_{n_z+1})&= \frac{1}{(n_x+1)(n_y+1)(n_z+1)}\sum_{i=1}^{n_x+1}\sum_{j=1}^{n_y+1}\sum_{k=1}^{n_z+1}I(x_i<t_{\min}^j<z_{k-1})\label{Lower3g}\\
\hspace{-3mm}\overline{P}(X_{n_x+1}<Y_{n_y+1}<Z_{n_z+1})&=\frac{1}{(n_x+1)(n_y+1)(n_z+1)}\sum_{i=1}^{n_x+1}\sum_{j=1}^{n_y+1}\sum_{k=1}^{n_z+1}I(x_{i-1}<t_{\max}^j<z_{k})\label{Upper3g}
\end{align}
where   $t_{\min}^j$ ($t_{\max}^j$ ) is any value belonging to a sub-interval (created by the $x$ and $z$ observations) within $(y_{j-1},y_j)$,  $j=1,\ldots,n_y+1$, such that the probability for the event $X_{n_x+1}<Y_{n_y+1}<Z_{n_z+1}$ is minimum (maximum). The proof  of these results is given in \cite{CMEC2014}, but a sketch  of how these lower and upper probabilities are obtained is provided below.


For the lower (upper) probability, the  probability mass $1/(n_x+1)$ corresponding to group $X$  will be assigned to the  right-end (left-end) of the intervals  $(x_{i-1},x_{i})$, $i=1,\ldots,n_x+1$,
while the probability mass $1/(n_z+1)$ corresponding to group $Z$ will be assigned to the left-end (right-end) of the intervals $(z_{k-1},z_{k})$,  $k=1,\ldots,n_z+1$. With regard to group $Y$,  the main question is how  the  probability  mass $1/(n_y+1)$ can be  assigned for each interval $(y_{j-1},y_j)$, $j=1,\ldots,n_y+1$.   Suppose there are  $n^{j}_x$ and $n^{j}_z$ observations from groups $X$ and $Z$ between $y_{j-1}$ and $y_{j}$, respectively. These observations create $n^{j}_x+n^{j}_z+1$ sub-intervals within $(y_{j-1},y_j)$.   Let $S_x^{t^j}$ ($S_z^{t^j}$) be the number of  assigned probability masses $1/(n_x+1)$  ($1/(n_z+1)$) to the left (right) of any value $t^j$ belonging to the $k^j$-th sub-interval  within $(y_{j-1},y_j)$, where $k^j=1,  \ldots, n^{j}_x+n^{j}_z+1$. 
Then   the lower (upper) probability can be obtained by minimising (maximising) the  quantity $K^j=S_x^{t^j}\times S_z^{t^j}$ over all these sub-intervals.  Let $t_{\min}^j$ ($t_{\max}^j$) be any value within the $k_{\min}^j$-th ($k_{\max}^j$-th) sub-interval that is  corresponding  to the minimum  (maximum) value of $K^j$. For the special case when  there are no observations from groups $X$ and $Z$ between $y_{j-1}$ and $y_{j}$,  the corresponding probability mass $1/(n_y+1)$ will be assigned to any value within this interval.\\


One may wish to avoid the optimisation process introduced above for large data sets and in particular when the groups are considerably overlapped, in which case obtaining easy to calculate bounds is attractive.  More discussion about the complexity of the optimisation process is given in Section \ref{sec.moregroups}. Below, we derive  lower and upper bounds for the NPI lower and upper probabilities \eqref{Lower3g} and \eqref{Upper3g}. First let us consider the lower bound for the lower probability, so  for $\underline{P}^L$ we required total separation for the intervals $(x_{i-1},x_i)$, $(y_{j-1},y_j)$ and $(z_{k-1},z_k)$, $i=1,\ldots,n_x+1$, $j=1,\ldots,n_y+1$  and $k=1,\ldots,n_z+1$, that is 
\begin{align}\label{LLprob}
\underline{P}^L&=\frac{1}{(n_x+1) (n_y+1) (n_z+1)}\sum_{i=1}^{n_x+1}\sum_{j=1}^{n_y+1}\sum_{k=1}^{n_z+1}I(x_{i}<y_{j-1} \boldsymbol{\wedge} y_{j}<z_{k-1})
\end{align}
and for the upper bound for the lower probability $\underline{P}^U$, the probability mass $1/(n_x+1)$ ($1/(n_z+1)$) corresponding to group $X$ ($Z$) will be assigned to the right-end (left-end) intervals created by the observations from this group. With regard to group $Y$ it does not matter whether we assign the probability mass $1/(n_y+1)$ to the right-end or left-end intervals created by the observations from  group $Y$, then \footnote{If the probability mass $1/(n_y+1)$ is assigned to the right-end of the intervals $(y_{j-1},y_j)$, $j=1,\ldots,n_y+1$,   that is  to the data $y$ observations and to $y_{n_y+1}=\infty$, then $\sum_{i=1}^{n_x+1}\sum_{k=1}^{n_z+1}I(x_i<y_{n_y+1}<z_{k-1})=0$ and thus $\sum_{i=1}^{n_x+1}\sum_{j=1}^{n_y+1}\sum_{k=1}^{n_z+1}I(x_i<y_j<z_{k-1})=\sum_{i=1}^{n_x+1}\sum_{j=1}^{n_y}\sum_{k=1}^{n_z+1}I(x_i<y_j<z_{k-1})$.  Similarly, if  the probability mass $1/(n_y+1)$  is assigned to the left-end of the intervals $(y_{j-1},y_j)$, $j=1,\ldots,n_y+1$, that is it will be assigned to $y_0=-\infty$ and to the data $y$ observations,  then $\sum_{i=1}^{n_x+1}\sum_{k=1}^{n_z+1}I(x_i<y_{0}<z_{k-1})=0$ and thus $\sum_{i=1}^{n_x+1}\sum_{j=1}^{n_y+1}\sum_{k=1}^{n_z+1}I(x_i<y_{j-1}<z_{k-1})=\sum_{i=1}^{n_x+1}\sum_{j=1}^{n_y}\sum_{k=1}^{n_z+1}I(x_i<y_j<z_{k-1})$.}

\begin{align}\label{ULprob}
&\underline{P}^U=\frac{1}{(n_x+1) (n_y+1) (n_z+1)}\sum_{i=1}^{n_x+1}\sum_{j=1}^{n_y+1}\sum_{k=1}^{n_z+1}I(x_i<y_j<z_{k-1})
\end{align}
Similarly,  for the lower bound for the upper probability $\overline{P}^L$, the probability mass $1/(n_x+1)$ ($1/(n_z+1)$) corresponding to group $X$ ($Z$) will be assigned to the left-end (right-end) intervals created by the observations from this group. With regard to group $Y$ it does not matter whether we assign the probability mass $1/(n_y+1)$ to the right-end or left-end intervals created by the observations from  group $Y$, then
\begin{align}\label{LUprob}
&\overline{P}^L=\frac{1}{(n_x+1) (n_y+1) (n_z+1)}\sum_{i=1}^{n_x+1}\sum_{j=1}^{n_y+1}\sum_{k=1}^{n_z+1}I(x_{i-1}<y_j<z_{k})
\end{align}
and for the upper bound for the upper probability $\overline{P}^U$, we count all possible combinations of the intervals $(x_{i-1},x_i)$, $(y_{j-1},y_j)$ and $(z_{k-1},z_k)$, $i=1,\ldots,n_x+1$, $j=1,\ldots,n_y+1$  and $k=1,\ldots,n_z+1$, for which we can find $x\in (x_{i-1},x_i)$, $y\in (y_{j-1},y_j)$ and $z\in (z_{k-1},z_k)$ such that  $x<y<z$, then 
\begin{align}\label{UUprob}
\overline{P}^U&=\frac{1}{(n_x+1) (n_y+1) (n_z+1)}\sum_{i=1}^{n_x+1}\sum_{j=1}^{n_y+1}\sum_{k=1}^{n_z+1}I(x_{i-1}<y_{j}\boldsymbol{\wedge} x_{i-1}<z_{k} \boldsymbol{\wedge} y_{j-1}<z_{k} )
\end{align}
The exact values of the lower and upper probabilities,  \eqref{Lower3g} and \eqref{Upper3g}, are nested between these corresponding 
 lower and upper bounds, that is
$\underline{P}^L\leq \underline{P} \leq \underline{P}^U$ and
$\overline{P}^L\leq \overline{P} \leq \overline{P}^U$.  Furthermore, 
the  exact  lower and upper probabilities  always bound the  empirical probability for the event $X<Y<Z$, which is given by 
\[\hat{H}=\frac{1}{n_x n_y n_z}\sum_{i=1}^{n_x}\sum_{j=1}^{n_y}\sum_{k=1}^{n_z}I(x_i<y_j<z_k).\]

 The results presented in this section have been used in \cite{CMEC2014} for  three-group ROC inference, but the inferences have not been considered for more than three groups. In this paper we extend these results to  more than three groups, which results in a more complex  optimisation process, and we also explore several applications including  multiple groups inference, ranked set sampling and assessment of diagnostic accuracy for more than three groups. But first we give an example for the three-group case.

\begin{example}\label{example1}{\rm
Suppose we have three groups, $X$, $Y$, and $Z$. Data of group $X$ are 2,  3, 5, 6, 7, 8, 10, 11, 15, 17, 18 and 21, for group $Y$ there are only two observations 9 and 20, and finally data of group $Z$ are 1,  4, 12, 13, 14, 16, 19, 22, 23, 24 and 25. These data are represented in Figure \ref{fig1}. The empirical probability   is $\hat{H}=98/264=0.3712$.

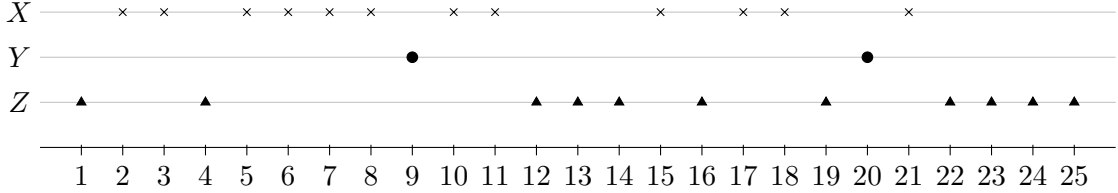
\begin{figure}
	\centering
\begin{tikzpicture}[y=.6cm, x=.55cm,font=\small]
\node[] at (-0.5,1) {$Z$};
\node[] at (-0.5,2) {$Y$};
\node[] at (-0.5,3) {$X$};
	\draw (0,0) -- coordinate (x axis mid) (26,0);
	\draw[lightgray,very thin] (0,1) --(26,1);
		\draw[lightgray,very thin] (0,2) --  (26,2);
	\draw[lightgray,very thin] (0,3) -- (26,3);
    	\foreach \x in {1,...,25}
     		\draw (\x,1pt) -- (\x,-3pt)
			node[anchor=north] {\x};
 		\draw plot[only marks, mark=x] 
	file {X.data};
		\draw plot[only marks, mark=*] 
	file {Y.data};
		\draw plot[only marks, mark=triangle*] 
	file {Z.data};
\end{tikzpicture}
\caption{Data set of Example \ref{example1}}
	\label{fig1}
\end{figure}

\begin{figure}
    \centering
    \includegraphics[scale=0.7]{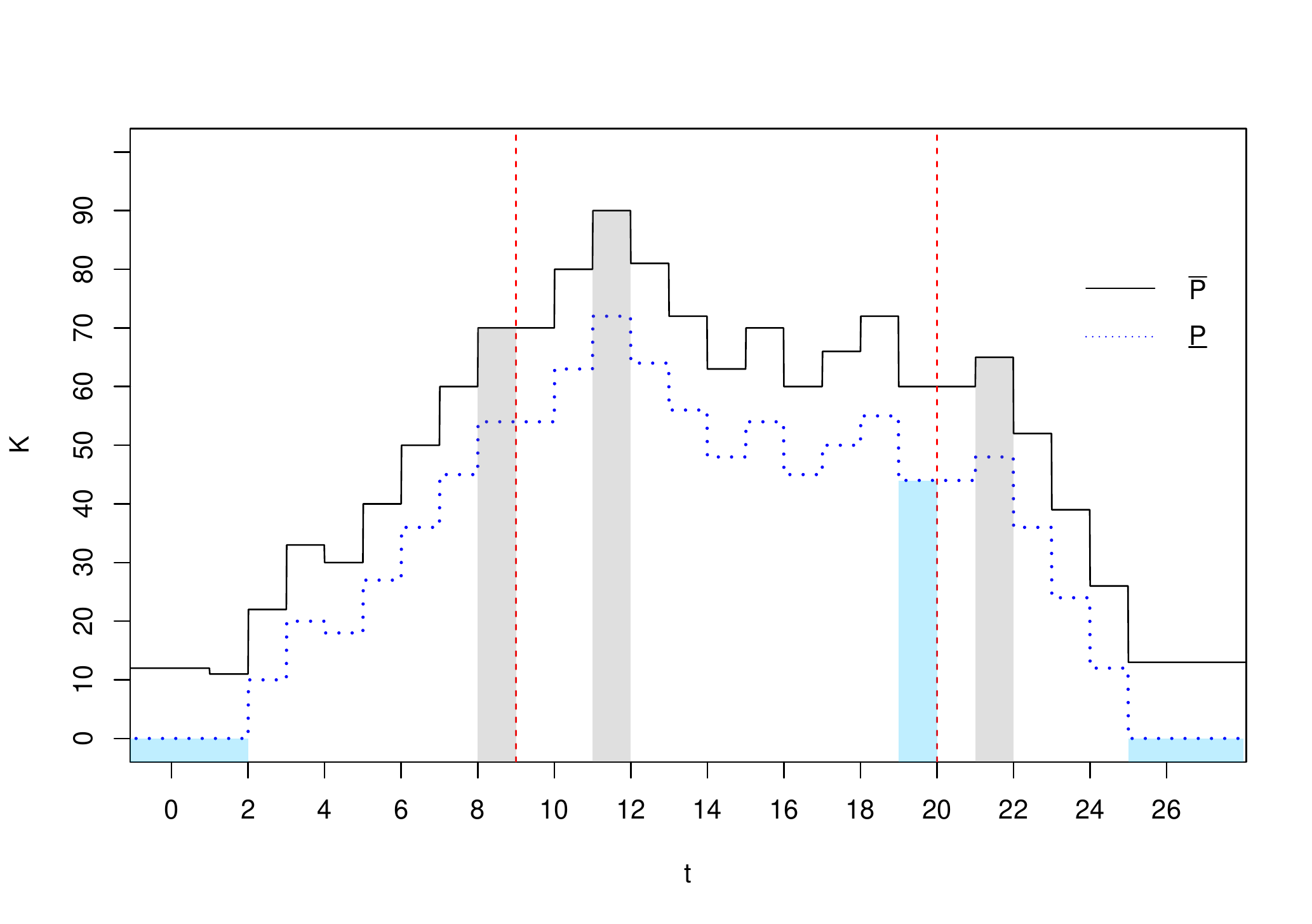}
    \caption{$K^j$ values over all sub-intervals, Example \ref{example1}}
    \label{fig1b}
\end{figure}

To get the lower probability, the   probability mass $\frac{1}{13}$ ($\frac{1}{12}$) corresponding to group $X$ ($Z$) will be assigned to the  right-end (left-end) intervals created by the observations of this group. However, to get the upper probability, the   probability mass $\frac{1}{13}$ ($\frac{1}{12}$) corresponding to group $X$ ($Z$) will be assigned to the  left-end (right-end) intervals created by the observations of this group. With regard to group $Y$, for the lower probability, the probability mass corresponding to $(-\infty,y_1)$, $(y_1,y_2)$ and $(y_2,\infty)$ will be assigned to the intervals $(-\infty,2)$, $(19,20)$ and $(25,\infty)$, respectively. For  the upper probability,  the probability mass corresponding to  these intervals will be assigned to the intervals $(8,9)$, $(11,12)$ and $(21,22)$, respectively.
Figure  \ref{fig1b} shows the $K^j$ values over all  sub-intervals, where the blue (grey) shaded areas are corresponding to the chosen sub-intervals that minimise (maximise) the quantity $K^j=S_x^{t^j}\times S_z^{t^j}$.

 The NPI lower and upper probabilities for the event  $X_{13}<Y_3<Z_{12}$ are  
\begin{align*}
\underline{P}&=\frac{1}{(13)(3)(12)}\sum_{j=1}^{3} S_x^{t_{\min}^j}\times S_z^{t_{\min}^j}=\frac{1}{(13)(3)(12)}(0+44+0)=\frac{44}{468}=0.0940\\
\overline{P}&=\frac{1}{(13)(3)(12)}\sum_{j=1}^{3}S_x^{t_{\max}^j}\times S_z^{t_{\max}^j}=\frac{1}{(13)(3)(12)}(70+90+65)=\frac{225}{468}=0.4808
\end{align*}
where  $t_{\min}^1\in(-\infty,2)$, $t_{\min}^2\in(19,20)$, $t_{\min}^3\in(25,\infty)$, $t_{\max}^1\in(8,9)$, $t_{\max}^2\in(11,12)$ and $t_{\max}^3\in(21,22)$. The lower and upper bounds for the lower probability, calculated from (\ref{LLprob}) and (\ref{ULprob}), are $\underline{P}^L=\frac{24}{468}=0.0513$ and $\underline{P}^U=\frac{98}{468}=0.2094$. The lower and upper bounds for the upper probability, calculated from (\ref{LUprob}) and (\ref{UUprob}), are $\overline{P}^L=\frac{130}{468}=0.2778$ and $\overline{P}^U=\frac{248}{468}=0.5299$. 


}
\end{example}

\subsection{The ordering of future observations for  $q>3$ groups }\label{sec.moregroups}

In this section we extend the results presented above for the situation when we have $q>3$ groups.  Suppose we have  $q>3$  independent   groups, $X_1, X_2, \ldots,X_q$, and the aim is to introduce  NPI lower and upper probabilities for the event $X_{1,n_1+1}<X_{2,n_2+1}<\ldots<X_{q,n_q+1}$, where $X_{j,n_j+1}$ is the next future observation from  group $j$, for $j=1,2,\ldots,q$. In order to introduce NPI for such an event we will apply $A_{(n)}$ per group, so we need to introduce the following notations. Let the ordered observations from group $j$  be denoted by $x_{j,1}<x_{j,2}<\ldots<x_{j,n_j}$ and let $x_{j,0}=-\infty$ and $x_{j,n_j+1}=\infty$ for ease of notation.  For $X_{j,n_j+1}$, representing a future observation from group $j$, $A_{(n_j)}$ partially specifies a probability distribution by 
$P(X_{j,n_j+1} \in (x_{j,i_j-1},x_{j,i_j})) = \frac{1}{n_j+1}$ for $i_j=1,\ldots,n_j+1$. The main aim is to find the NPI lower and upper probabilities for the event $X_{1,n_1+1}<X_{2,n_2+1}<\ldots<X_{q,n_q+1}$, that is
\begin{equation}\label{EXACTL}
\underline{P}(X_{1,n_1+1}<X_{2,n_2+1}<\ldots<X_{q,n_q+1})
\end{equation}
\begin{equation}\label{EXACTU}
\overline{P}(X_{1,n_1+1}<X_{2,n_2+1}<\ldots<X_{q,n_q+1})
\end{equation}
which in turn bound  
the empirical value for the event $X_1<X_2<\ldots<X_q$ given as
\[\hat{H}=\frac{1}{n_1n_2\ldots n_q}\sum_{i_1=1}^{n_1}\sum_{i_2=1}^{n_2}\ldots \sum_{i_q=1}^{n_q}I(x_{1,i_1}<x_{2,i_2}<\ldots<x_{q,i_q}).\]

While it is clear how  the probability masses $1/(n_1+1)$ and $1/(n_q+1)$  corresponding to the first and the last group can be assigned, respectively,  we need to find  for the remaining groups  how the probability mass $1/(n_j+1)$ should be assigned for each interval $(x_{j,i_{j-1}},x_{j,i_j})$, $j=2,\ldots,q-1$ and $i_j=1,\ldots,n_j+1$, taking into account all observations from the other groups. This  has to be done simultaneously in order to minimise for the lower probability and to maximise for the upper probability. Therefore, first we are going to introduce lower and upper bounds for these lower and upper probabilities as we did in the previous section, then we are going to propose two alternative approximations of them.

\subsubsection{Lower and upper bounds for the NPI lower and upper probabilities}
In order to find the lower bound for the lower probability  $\underline{P}^L$ we require total separation for the intervals $(x_{j,i_j-1},x_{j,i_j})$, $i_j=1,2,\ldots,n_{j}+1$ and $j=1,2,\ldots,q$, thus
\begin{align}\label{LLprob.q}
\underline{P}^L&=\frac{1}{\prod_{j=1}^{q}(n_j+1)}\sum_{i_1=1}^{n_1+1}\sum_{i_2=1}^{n_2+1}\ldots \sum_{i_q=1}^{n_q+1} \prod_{j=1}^{q-1} I(x_{j,i_j}<x_{j+1,i_{j+1}-1})
\end{align}
For the upper bound for the lower probability $\underline{P}^U$, the probability mass $1/(n_1+1)$ ($1/(n_q+1)$) corresponding to group $X_1$ ($X_q$) will be assigned to the right-end (left-end) intervals created by the observations from this group, that is $(x_{1,i_1-1},x_{1,i_1})$ ($(x_{q,i_q-1},x_{q,i_q})$). With regard to groups $X_2,\ldots,X_{q-1}$ it does not matter whether we assign the probability mass corresponding to these groups to the right-end or left-end intervals created by the observations from  these groups, then
\begin{align}\label{ULprob.q}
&\underline{P}^U=\frac{1}{\prod_{j=1}^{q}(n_j+1)}\sum_{i_1=1}^{n_1+1}\sum_{i_2=1}^{n_2+1}\ldots\sum_{i_{q-1}=1}^{n_{q-1}+1} \sum_{i_q=1}^{n_q+1}I(x_{1,i_1}<x_{2,i_2}<\ldots<x_{q-1,i_{q-1}}<x_{q,i_q-1})
\end{align}
Similarly,  for the lower bound of the upper probability $\overline{P}^L$, the probability mass $1/(n_1+1)$ ($1/(n_q+1)$) corresponding to group $X_1$ ($X_q$) will be assigned to the left-end (right-end) intervals created by the observations from this group, that is $(x_{1,i_1-1},x_{1,i_1})$ ($(x_{q,i_q-1},x_{q,i_q})$). With regard to groups $X_2,\ldots,X_{q-1}$ it does not matter whether we assign the probability mass corresponding to these groups to the right-end or left-end intervals created by the observations from  these groups, then
\begin{align}\label{LUprob.q}
&\overline{P}^L=\frac{1}{\prod_{j=1}^{q}(n_j+1)}\sum_{i_1=1}^{n_1+1}\sum_{i_2=1}^{n_2+1}\ldots\sum_{i_{q-1}=1}^{n_{q-1}+1} \sum_{i_q=1}^{n_q+1}I(x_{1,i_1-1}<x_{2,i_2}<\ldots<x_{q-1,i_{q-1}}<x_{q,i_q})
\end{align}
Finally  for the upper bound for the upper probability $\overline{P}^U$ we count all combinations of the intervals $(x_{j,i_j-1},x_{j,i_j})$, $i_j=1,\ldots,n_{j}+1$ and $j=1,\ldots,q$, for which we can find any values within these intervals such that $x_1<x_2<\ldots<x_q$, which leads to
\begin{align}\label{UUprob.q}
\overline{P}^U&=\frac{1}{\prod_{j=1}^{q}(n_j+1)}\sum_{i_1=1}^{n_1+1}\sum_{i_2=1}^{n_2+1}\ldots \sum_{i_q=1}^{n_q+1} \prod_{j=1}^{q-1}\prod_{k=j+1}^{q} I(x_{j,i_j-1}<x_{k,i_{k}})
\end{align}

If all observations are perfectly ordered, meaning that all observations from group $X_1$ are less than all observations from group $X_2$, and so on until  all observations from group $X_{q-1}$ are less than all observations from group $X_q$. Then the probabilities, \eqref{LLprob.q}, \eqref{ULprob.q}, \eqref{LUprob.q} and \eqref{UUprob.q}, reduced to
\begin{align}
\underline{P}^{L*}&=\frac{1}{\prod_{i=1}^q(n_i+1)} n_1 n_q \prod_{i=2}^{q-1}(n_i-1) \label{eq.per1}\\
\underline{P}^{U*}&=\frac{1}{\prod_{i=1}^q(n_i+1)}\prod_{i=1}^{q}n_i \label{eq.per2}\\
\overline{P}^{L*}&=\frac{1}{\prod_{i=1}^q(n_i+1)} (n_1+1) (n_q+1)\prod_{i=2}^{q-1}n_i \label{eq.per3}\\
\overline{P}^{U*}&=\frac{1}{\prod_{i=1}^q(n_i+1)}\prod_{i=1}^q(n_i+1)=1.\label{eq.per4}
\end{align}

These probabilities will be used in Section \ref{sec.app.1} to see how far the data observations are from the perfect ordering case.

\vspace{.3cm}

\subsubsection{Exact NPI lower and upper probabilities}
The main challenge  is to find the exact lower and upper probabilities in \eqref{EXACTL} and \eqref{EXACTU}, respectively. For the lower probability $\underline{P}$, and in order to minimise the probability for the event  $X_{1,n_1+1}<X_{2,n_2+1}<\ldots<X_{q,n_q+1}$, the probability mass $1/(n_1+1)$ ($1/(n_q+1)$) corresponding to group $X_1$ ($X_q$) will be assigned to the right-end (left-end) intervals created by the observations from this group, that is $(x_{1,i_1-1},x_{1,i_1})$ ($(x_{q,i_q-1},x_{q,i_q})$), then
\begin{align}\label{Lprob.q}
&\underline{P}=\frac{1}{\prod_{j=1}^{q}(n_j+1)}\sum_{i_1=1}^{n_1+1}\sum_{i_2=1}^{n_2+1}\ldots\sum_{i_{q-1}=1}^{n_{q-1}+1} \sum_{i_q=1}^{n_q+1}P(x_{1,i_1}<X_{2,n_2+1}<\ldots<X_{q-1,n_{q-1}+1}<x_{q,i_q-1}|C_1)
\end{align}
where $C_1=\{X_{j,n_j+1}\in(x_{j,i_j-1},x_{j,i_j}), j=2,\ldots,q-1\}$.  For
  the upper probability $\overline{P}$,   in order to maximise the probability for the event  $X_{1,n_1+1}<X_{2,n_2+1}<\ldots<X_{q,n_q+1}$, the probability mass $1/(n_1+1)$ ($1/(n_q+1)$) corresponding to group $X_1$ ($X_q$) will be assigned to the left-end (right-end) intervals created by the observations from this group, that is $(x_{1,i_1-1},x_{1,i_1})$ ($(x_{q,i_q-1},x_{q,i_q})$), then
\begin{align}\label{Uprob.q}
&\overline{P}=\frac{1}{\prod_{j=1}^{q}(n_j+1)}\sum_{i_1=1}^{n_1+1}\sum_{i_2=1}^{n_2+1}\ldots\sum_{i_{q-1}=1}^{n_{q-1}+1} \sum_{i_q=1}^{n_q+1}P(x_{1,i_1-1}<X_{2,n_2+1}<\ldots<X_{q-1,n_{q-1}+1}<x_{q,i_q}|C_1)
\end{align}
with  $C_1$ as above.  This is similar to the argument presented by \citet{CF96} for two groups.\\

As discussed earlier, if one wishes to obtain these exact lower and upper probabilities,   we need to find,  for the remaining groups,  how the probability masses $1/(n_j+1)$ should be distributed over each interval $(x_{j,i_{j-1}},x_{j,i_j})$, $j=2,\ldots,q-1$ and $i_j=1,\ldots,n_j+1$, taking into account all observations from the other groups.  Suppose there are $n^{j,i_j}_l$ observations from groups $X_l$, $l=1,\ldots, q$ and $l\neq j$ between $x_{j,i_{j-1}}$ and $x_{j,i_j}$. These observations create $\sum_{l\neq j}n^{j,i_j}_l+1$ sub-intervals  within $(x_{j,i_{j-1}},x_{j,i_j})$, denoted by $I^{j,i_j}_{k_{j,i_j}}$, $k_{j,i_j}=1,2,\ldots,\sum_{l\neq j}n^{j,i_j}_l+1$. We then follow a similar procedure to the three-group case, however this  has to be done simultaneously, for the remaining $q-2$ groups,  in order to minimise for the lower probability and to maximise for the upper probability.

 Therefore, for more than three groups it may become computationally cumbersome to find the exact lower and upper probabilities, in particular when we have large data sets and  the groups are considerably overlapped. Explicitly, the number of ways  the probability masses can be assigned, simultaneously, for the remaining $q-2$ groups,  is 
\[  \prod_{j=2}^{q-1} \prod_{i_j=1}^{n_j+1} \left(\sum_{l\neq j}n^{j,i_j}_l+1\right)\]
which reduces to 
\[  \prod_{j=2}^{q-1} \left(\sum_{l<j}n_l+1\right) \left(\sum_{l> j}n_l+1\right)\]
when the groups are fully separated, which is corresponding to the number of ways  the probability masses can be assigned  for the first and last intervals for group $j=2,\ldots,q-2$.


\subsubsection{Two heuristic algorithms} 
Below we introduce  two heuristic algorithms to find  approximations for the above NPI lower and upper probabilities.\\
  
\noindent {\bf Algorithm A:} One way of  finding a reasonable approximation for the lower and upper probabilities in \eqref{EXACTL} and \eqref{EXACTU}, is    to perform an optimisation  over one group, $j=2,\ldots,q-1$, each time and then take the minimum of these values to approximate the lower probability and   the maximum of these values to approximate the  upper probability. The minimum (maximum) of these lower (upper) probabilities will be the approximation for the exact lower (upper) probability $\underline{P}$ ($\overline{P}$). We can summarise the algorithm in the following steps:

\begin{enumerate}
	\item Consider one group $j=2,\ldots,q-1$, say group 2, and  optimise (minimise for the lower and  maximise for the upper) over this group.
    \item For the other groups, the corresponding probability masses are assigned to either the right-end or to the left-end intervals created by the observations per group.
	\item Calculate the lower and upper probabilities using equations  (\ref{Lprob.q}) and (\ref{Uprob.q}).
	\item Repeat steps 1-3, for all $j=3,\ldots,q-1$.
	\item Take the minimum (maximum) of these lower (upper) probabilities which we use as reasonable approximations for the exact lower (upper) probability $\underline{P}$ ($\overline{P}$).\\
\end{enumerate}

\noindent {\bf Algorithm B:} The same as Algorithm A, but the last step is replaced by the following step:
\begin{enumerate}
	\item[5.]  We combine the optimisation values resulted from steps 1-4,  such that for the overlapping intervals corresponding to groups $j=2,\ldots,q-1$,   the   probability masses associated with these intervals are appropriately assigned in order to minimise for the lower probability and to maximise for the upper probability.
\end{enumerate}

To elaborate, as the result of the optimisation process in steps 1-4,  the probability mass $1/(n_j+1)$  is   assigned to a sub-interval $I^{j,i_j}_{k^*_{j,i_j}}$ within $(x_{j,i_j-1}, x_{j,i_j})$, for $j=2,\ldots,q-1$,  $i_j=1,\ldots,n_j+1$, and $k^*_{j,i_j} \in \{1,2,\ldots, \sum_{l\neq j} n^{j,i_j}_{l}+1\}$. So in algorithm B, we take all these sub-intervals  $I^{j,i_j}_{k^*_{j,i_j}}$ and simultaneously try to assign the corresponding probability masses to values within these sub-intervals in order to minimise for the lower and to maximise for the upper. So  algorithm B could be computationally expensive if the groups  considerably overlap  each other. 


The NPI lower and upper bounds and the two  heuristic algorithms provide approximations to the exact NPI lower and upper probabilities, this is  illustrated in the following example. The R code and data sets for the examples in this article are available from the author's website. \\[2ex]



\begin{example}\label{example2}{\rm
In this example we use the data set  presented   in Figure \ref{fig6}  to illustrate the proposed method. Here we have four groups $q=4$ where $n_1=4$, $n_2=5$, $n_3=5$, and $n_4=6$. The empirical value for the event $X_1<X_2<X_3<X_4$ is 
\[\hat{H}=\frac{1}{n_1n_2n_3n_4}\sum_{i_1=1}^{n_1}\sum_{i_2=1}^{n_2}\sum_{i_3=1}^{n_3}\sum_{i_4=1}^{n_4}I(x_{1,i_1}<x_{2,i_2}<x_{3,i_3}<x_{4,i_4})=\frac{47}{600}=0.0783\]

\begin{figure}
	\centering
\begin{tikzpicture}[y=.6cm, x=.55cm,font=\small]
\node[] at (-0.5,1) {$X_4$};
\node[] at (-0.5,2) {$X_3$};
\node[] at (-0.5,3) {$X_2$};
\node[] at (-0.5,4) {$X_1$};
	\draw (0,0) -- coordinate (x axis mid) (21,0);
	\draw[lightgray,very thin] (0,1) --(21,1);
		\draw[lightgray,very thin] (0,2) --  (21,2);
	\draw[lightgray,very thin] (0,3) -- (21,3);
	\draw[lightgray,very thin] (0,4) --  (21,4);
    	\foreach \x in {1,...,20}
     		\draw (\x,1pt) -- (\x,-3pt)
			node[anchor=north] {\x};
 		\draw plot[only marks, mark=x] 
	file {X1.data};
		\draw plot[only marks, mark=*] 
	file {X2.data};
		\draw plot[only marks, mark=square*] 
	file {X3.data};
		\draw plot[only marks, mark=triangle*] 
	file {X4.data};
\end{tikzpicture}
	\caption{Data set of Example \ref{example2}}
	\label{fig6}
\end{figure}
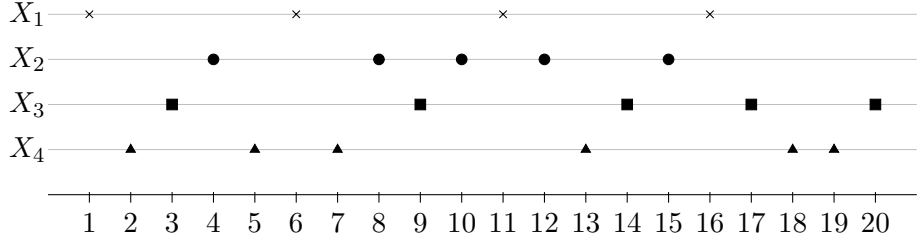

The  lower and upper bounds, given in equations \eqref{LLprob.q}, \eqref{ULprob.q}, \eqref{LUprob.q} and \eqref{UUprob.q}, for the NPI lower and upper probabilities for the event $X_{1,n_1+1}<X_{2,n_2+1}<X_{3,n_3+1}<X_{4,n_4+1}$ are
\begin{align*}
\underline{P}^L&=\frac{1}{\prod_{j=1}^{4}(n_j+1)}\sum_{i_1=1}^{n_1+1}\sum_{i_2=1}^{n_2+1}\sum_{i_3=1}^{n_3+1} \sum_{i_4=1}^{n_4+1} I(x_{1,i_1}<x_{2,i_{2}-1}) I(x_{2,i_2}<x_{3,i_{3}-1})I(x_{3,i_3}<x_{4,i_{4}-1})\\
&=\frac{12}{1260}=0.0095\\
\underline{P}^U&=\frac{1}{\prod_{j=1}^{4}(n_j+1)}\sum_{i_1=1}^{n_1+1}\sum_{i_2=1}^{n_2+1}\sum_{i_{3}=1}^{n_{3}+1} \sum_{i_4=1}^{n_4+1}I(x_{1,i_1}<x_{2,i_2}<x_{3,i_3}<x_{4,i_4-1})\\
&=\frac{47}{1260}=0.0373
\end{align*}
and 
\begin{align*}
\overline{P}^L&=\frac{1}{\prod_{j=1}^{4}(n_j+1)}\sum_{i_1=1}^{n_1+1}\sum_{i_2=1}^{n_2+1}\sum_{i_{3}=1}^{n_{3}+1} \sum_{i_4=1}^{n_4+1}I(x_{1,i_1-1}<x_{2,i_2}<x_{3,i_{3}}<x_{4,i_4})\\
&=\frac{120}{1260}=0.0952\\
\overline{P}^U&=\frac{1}{\prod_{j=1}^{4}(n_j+1)}\sum_{i_1=1}^{n_1+1}\sum_{i_2=1}^{n_2+1}\sum_{i_3=1}^{n_3+1} \sum_{i_4=1}^{n_4+1}I(x_{1,i_1-1}<x_{2,i_{2}})I(x_{1,i_1-1}<x_{3,i_{3}})I(x_{1,i_1-1}<x_{4,i_{4}}) \\
&\hspace{6.2cm}I(x_{2,i_2-1}<x_{3,i_{3}})I(x_{2,i_2-1}<x_{4,i_{4}})I(x_{3,i_3-1}<x_{4,i_{4}})\\
&=\frac{266}{1260}=0.2111
\end{align*}

\noindent  Next, we  implement algorithms A and B to find approximations for the following   lower and upper probabilities,
\begin{align*}
&\underline{P}=\frac{1}{\prod_{j=1}^{4}(n_j+1)}\sum_{i_1=1}^{n_1+1}\sum_{i_2=1}^{n_2+1}\sum_{i_{3}=1}^{n_{3}+1} \sum_{i_4=1}^{n_4+1}P(x_{1,i_1}<X_{2,n_2+1}<X_{3,n_{3}+1}<x_{4,i_4-1}|C_1)
\end{align*}
\begin{align*}
&\overline{P}=\frac{1}{\prod_{j=1}^{4}(n_j+1)}\sum_{i_1=1}^{n_1+1}\sum_{i_2=1}^{n_2+1}\sum_{i_{3}=1}^{n_{3}+1} \sum_{i_4=1}^{n_4+1}P(x_{1,i_1-1}<X_{2,n_2+1}<X_{3,n_{3}+1}<x_{4,i_4}|C_1)
\end{align*}
where $C_1=\{X_{2,n_2+1}\in(x_{2,i_2-1},x_{2,i_2}), X_{3,n_3+1}\in(x_{3,i_3-1},x_{3,i_3})\}$.\\

First we are going to optimise over  group 2, that is to minimise for the lower probability and maximise for the upper probability, thus
\begin{align}\label{ex2.opt.2L}
&\underline{P}=\frac{1}{\prod_{j=1}^{4}(n_j+1)}\sum_{i_1=1}^{n_1+1}\sum_{i_2=1}^{n_2+1}\sum_{i_{3}=1}^{n_{3}+1} \sum_{i_4=1}^{n_4+1}P(x_{1,i_1}<X_{2,n_2+1}<x_{3,i_3}<x_{4,i_4-1}|X_{2,n_2+1}\in(x_{2,i_2-1},x_{2,i_2}))
\end{align}
\begin{align}\label{ex2.opt.2U}
&\overline{P}=\frac{1}{\prod_{j=1}^{4}(n_j+1)}\sum_{i_1=1}^{n_1+1}\sum_{i_2=1}^{n_2+1}\sum_{i_{3}=1}^{n_{3}+1} \sum_{i_4=1}^{n_4+1}P(x_{1,i_1-1}<X_{2,n_2+1}<x_{3,i_3}<x_{4,i_4}|X_{2,n_2+1}\in(x_{2,i_2-1},x_{2,i_2}))
\end{align}
where  for group 3,   it does not matter whether the  probabilities masses are assigned to the right-end or to the left-end intervals created by the observations from this group.

In order to minimise (\ref{ex2.opt.2L}) over group 2, we find that the  probability masses corresponding to $X_{2,n_2+1}$ should be assigned to the intervals $(-\infty,1)$, $(4,6)$, $(9,10)$, $(10,11)$, $(14,15)$ and $(17,\infty)$. That is from (\ref{ex2.opt.2L}),
\begin{align*}
\underline{P}^{min2}&=\frac{1}{1260}\sum_{i_1=1}^{5}\sum_{i_{3}=1}^{5} \sum_{i_4=1}^{7}\left\{I(x_{1,i_1}<x_{2,1}<x_{3,i_3}<x_{4,i_4-1}|x_{2,1}\in (-\infty,1))\right.\\
&\hspace{4cm}+I(x_{1,i_1}<x_{2,2}<x_{3,i_3}<x_{4,i_4-1}|x_{2,2}\in (4,6))\\
&\hspace{4cm}+I(x_{1,i_1}<x_{2,3}<x_{3,i_3}<x_{4,i_4-1}|x_{2,3}\in (9,10))\\
&\hspace{4cm}+I(x_{1,i_1}<x_{2,4}<x_{3,i_3}<x_{4,i_4-1}|x_{2,4}\in (10,11))\\
&\hspace{4cm}+I(x_{1,i_1}<x_{2,5}<x_{3,i_3}<x_{4,i_4-1}|x_{2,5}\in (14,15))\\
&\hspace{4cm}\left.+I(x_{1,i_1}<x_{2,6}<x_{3,i_3}<x_{4,i_4-1}|x_{2,6}\in (17,\infty))\right\}\\
&= \frac{1}{1260} \{0+7+8+8+6+0\}=\frac{29}{1260} =0.0230
\end{align*}

Similarly for the upper probability, in order to maximise (\ref{ex2.opt.2U}) over group 2, we find that the  probability masses corresponding to $X_{2,n_2+1}$ should be assigned to the intervals $(1,3)$, $(6,8)$, $(8,9)$, $(11,12)$, $(12,14)$ and $(16,17)$. That is from (\ref{ex2.opt.2U}),
\begin{align*}
\overline{P}^{max2}&=\frac{1}{1260}\sum_{i_1=1}^{5}\sum_{i_{3}=1}^{5} \sum_{i_4=1}^{7}\left\{I(x_{1,i_1-1}<x_{2,1}<x_{3,i_3}<x_{4,i_4}|x_{2,1}\in (1,3))\right.\\
&\hspace{4cm}+I(x_{1,i_1-1}<x_{2,2}<x_{3,i_3}<x_{4,i_4}|x_{2,2}\in (6,8))\\
&\hspace{4cm}+I(x_{1,i_1-1}<x_{2,3}<x_{3,i_3}<x_{4,i_4}|x_{2,3}\in (8,9))\\
&\hspace{4cm}+I(x_{1,i_1-1}<x_{2,4}<x_{3,i_3}<x_{4,i_4}|x_{2,4}\in (11,12))\\
&\hspace{4cm}+I(x_{1,i_1-1}<x_{2,5}<x_{3,i_3}<x_{4,i_4}|x_{2,5}\in (12,14))\\
&\hspace{4cm}\left.+I(x_{1,i_1-1}<x_{2,6}<x_{3,i_3}<x_{4,i_4}|x_{2,6}\in (16,17))\right\}\\
&= \frac{1}{1260} \{34+33+33+28+28+20\}=\frac{176}{1260} =0.1397
\end{align*}

Secondly, we optimise over  group 3, so we minimise for the lower probability and maximise for the upper probability, where  for group 2    we are indifferent between assigning the corresponding probability masses  to the right-end or to the left-end intervals created by the observations from this group, thus

\begin{align}\label{ex2.opt.3L}
&\underline{P}=\frac{1}{\prod_{j=1}^{4}(n_j+1)}\sum_{i_1=1}^{n_1+1}\sum_{i_2=1}^{n_2+1}\sum_{i_{3}=1}^{n_{3}+1} \sum_{i_4=1}^{n_4+1}P(x_{1,i_1}<x_{2,i_2}<X_{3,n_3+1}<x_{4,i_4-1}|X_{3,n_3+1}\in(x_{3,i_3-1},x_{3,i_3}))
\end{align}
\begin{align}\label{ex2.opt.3U}
&\overline{P}=\frac{1}{\prod_{j=1}^{4}(n_j+1)}\sum_{i_1=1}^{n_1+1}\sum_{i_2=1}^{n_2+1}\sum_{i_{3}=1}^{n_{3}+1} \sum_{i_4=1}^{n_4+1}P(x_{1,i_1-1}<x_{2,i_2}<X_{3,n_3+1}<x_{4,i_4}|X_{3,n_3+1}\in(x_{3,i_3-1},x_{3,i_3}))
\end{align}

In order to minimise (\ref{ex2.opt.3L}) over group 3, we find that the  probability masses corresponding to $X_{3,n_3+1}$ should be assigned to the intervals $(-\infty,3)$, $(3,4)$, $(9,10)$,  $(14,15)$, $(19,20)$ and $(20,\infty)$. Then (\ref{ex2.opt.3L}) becomes,

\begin{align*}
\underline{P}^{min3}&=\frac{1}{1260}\sum_{i_1=1}^{5}\sum_{i_{2}=1}^{5} \sum_{i_4=1}^{7}\left\{I(x_{1,i_1}<x_{2,i_2}<x_{3,1}<x_{4,i_4-1}|x_{3,1}\in (-\infty,3))\right.\\
&\hspace{4cm}+I(x_{1,i_1}<x_{2,i_2}<x_{3,2}<x_{4,i_4-1}|x_{3,2}\in (3,4))\\
&\hspace{4cm}+I(x_{1,i_1}<x_{2,i_2}<x_{3,3}<x_{4,i_4-1}|x_{3,3}\in (9,10))\\
&\hspace{4cm}+I(x_{1,i_1}<x_{2,i_2}<x_{3,4}<x_{4,i_4-1}|x_{3,4}\in (14,15))\\
&\hspace{4cm}+I(x_{1,i_1}<x_{2,i_2}<x_{3,5}<x_{4,i_4-1}|x_{3,5}\in (19,20))\\
&\hspace{4cm}\left.+I(x_{1,i_1}<x_{2,i_2}<x_{3,6}<x_{4,i_4-1}|x_{3,6}\in (20,\infty))\right\}\\
&= \frac{1}{1260} \{0+0+9+16+0+0\}=\frac{25}{1260} =0.0198
\end{align*}


In order to maximise (\ref{ex2.opt.3U}) over group 3, we find that the  probability masses corresponding to $X_{3,n_3+1}$ should be assigned to the intervals $(-\infty,3)$, $(8,9)$, $(12,13)$,  $(15,17)$, $(17,18)$ and $(20,\infty)$. Then (\ref{ex2.opt.3U}) becomes,

\begin{align*}
\overline{P}^{max3}&=\frac{1}{1260}\sum_{i_1=1}^{5}\sum_{i_{2}=1}^{5} \sum_{i_4=1}^{7}\left\{I(x_{1,i_1-1}<x_{2,i_2}<x_{3,1}<x_{4,i_4}|x_{3,1}\in (-\infty,3))\right.\\
&\hspace{4cm}+I(x_{1,i_1-1}<x_{2,i_2}<x_{3,2}<x_{4,i_4}|x_{3,2}\in (8,9))\\
&\hspace{4cm}+I(x_{1,i_1-1}<x_{2,i_2}<x_{3,3}<x_{4,i_4}|x_{3,3}\in (12,13))\\
&\hspace{4cm}+I(x_{1,i_1-1}<x_{2,i_2}<x_{3,4}<x_{4,i_4}|x_{3,4}\in (15,17))\\
&\hspace{4cm}+I(x_{1,i_1-1}<x_{2,i_2}<x_{3,5}<x_{4,i_4}|x_{3,5}\in (17,18))\\
&\hspace{4cm}\left.+I(x_{1,i_1-1}<x_{2,i_2}<x_{3,6}<x_{4,i_4}|x_{3,6}\in (20,\infty))\right\}\\
&= \frac{1}{1260} \{0+20+48+48+48+16\}=\frac{180}{1260} =0.1429
\end{align*}

Now according to algorithm A, we can take the minimum of the two values $\underline{P}^{min2}$ and $\underline{P}^{min3}$ as an approximation for the lower probability, i.e.\ $\underline{P}^A= \frac{25}{1260} =0.0198$. And
we take the maximum of  $\overline{P}^{max2}$ and $\overline{P}^{max3}$ as an approximation for the upper probability, i.e.\ $\overline{P}^A= \frac{180}{1260} =0.1429$.

Using algorithm B, we can  improve  that further by combining these  optimisation processes  together such that for the overlapping intervals corresponding to $X_{2,n_2+1}$ and $X_{3,n_3+1}$ we appropriately assign the   probability masses associated with these intervals to minimise the lower probability further. For example we choose the value of $x_{3,1}$ within the interval $(-\infty,1)$ and to the left of $x_{2,1}$, and 
 we choose the value of $x_{3,3}$ within the interval $(9,10)$ and to the left of $x_{2,3}$, and we choose the value of $x_{3,4}$ within the interval $(14,15)$ and to the left of $x_{2,5}$, and finally we choose the value of $x_{2,6}$ within $(17,\infty)$ to be within any intervals beyond  20. Say, $x^*_{2,i_2}=\{0.5,5.5,9.5,10.5,14.5,20.5\}$ and 
$x^*_{3,i_3}=\{0.2,3.2,9.2,14.2,19.2,20.2\}$ then
 \begin{align*}
\underline{P}^B&=\frac{1}{1260}\sum_{i_1=1}^{5}\sum_{i_2=1}^{6}\sum_{i_3=1}^{6} \sum_{i_4=1}^{7}I(x_{1,i_1}<x^*_{2,i_2}<x^*_{3,i_3}<x_{4,i_4-1})=\frac{13}{1260} =0.0103
\end{align*}

 Similarly, the approximation for the upper probability can be obtained. That is, for the overlapping intervals corresponding to $X_{2,n_2+1}$ and $X_{3,n_3+1}$ we appropriately assign the   probability masses associated with these intervals to maximise the upper probability further. To achieve that we choose the value of $x_{3,1}$ within the interval $(1,2)$ and to the right of $x_{2,1}$, and we choose the value of $x_{3,2}$ within the interval $(8,9)$ and to the right of $x_{2,3}$, and we choose the value of $x_{3,3}$ within the interval $(12,13)$ and to the right of $x_{2,5}$, and finally we choose the value of $x_{3,4}$ within the interval $(16,17)$ and to the right of $x_{2,6}$. Say, $x^*_{2,i_2}=\{1.2,6.2,8.2,11.2,12.2,16.2\}$ and 
$x^*_{3,i_3}=\{1.5,8.5,12.5,16.5,17.5,20.5\}$ then
 \begin{align*}
\overline{P}^B&=\frac{1}{1260}\sum_{i_1=1}^{5}\sum_{i_2=1}^{6}\sum_{i_3=1}^{6} \sum_{i_4=1}^{7}I(x_{1,i_1-1}<x^*_{2,i_2}<x^*_{3,i_3}<x_{4,i_4})
=\frac{257}{1260} =0.2040
\end{align*}

To sum up, in order to  find approximations for the exact  lower and upper probabilities, we first optimise over one group (not the first or the last group) each time, and then we considered two algorithms: In algorithm A  we take the maximum for the upper and  the minimum for the lower of the values resulting from this stage  as an approximation for the exact lower and upper probabilities. Algorithm B uses the optimisation values and runs another optimisation process between these groups (again not the first or the last group), in our example we have only two groups $X_2$ and $X_3$ so we optimise again over these two groups together. As we can see,  the  lower (upper) probability obtained from algorithm B  is  very  close to the lower (upper) bound for the lower (upper) probability $\underline{P}^L$ ($\overline{P}^U$). However, applying algorithm A give us values for the lower and upper probabilities falls almost in the middle between its lower and upper bounds. 
  In fact, the exact lower and upper probabilities for this example are equal to the results given by algorithm B. But this is not necessarily always the case, as in algorithm B, we optimise over one group at a time while for the exact probabilities the optimisation processes are performed  simultaneously. We expect algorithm B to be closer to the exact results but more expensive computationally compared to algorithm A. Hereafter, only the results obtained from algorithm A  will be reported.

We can also consider different orderings of the groups, for example if we search among all the combinations of the order of ($X_1,X_2,X_3,X_4$)  that give the minimum and the maximum of $\hat{H}$ for this data set, we found these   are $X_2<X_3<X_4<X_1$ and $X_1<X_4<X_2<X_3$, where  $\hat{H}=\frac{2}{600}=0.0033$ and $\hat{H}=\frac{58}{600}=0.0967$, respectively.  Similarly for the event $X_{1,n_1+1}<X_{4,n_4+1}<X_{2,n_2+1}<X_{3,n_3+1}$, the  lower and upper bounds for the NPI lower and upper probabilities for this event are $\underline{P}^L=\frac{16}{1260}= 0.0127$, $\underline{P}^U=\frac{58}{1260}=0.0460$, $\overline{P}^L=\frac{134}{1260}=0.1063$ and $\overline{P}^U=\frac{296}{1260}=0.2349$. And the  lower and upper bounds for the NPI lower and upper probabilities for the event  $X_{2,n_2+1}<X_{3,n_3+1}<X_{4,n_4+1}<X_{1,n_1+1}$ are $\underline{P}^L=\frac{0}{1260}=0 $, $\underline{P}^U=\frac{2}{1260}=0.0016$, $\overline{P}^L=\frac{45}{1260}=0.0357$ and $\overline{P}^U=\frac{139}{1260}=0.1103$. We can similarly as above use the two  heuristic techniques  to approximate the exact NPI lower and upper probabilities.\\

}
\end{example}

\section{Some applications} \label{sec.app}
In this section, we show via examples how  the NPI lower and upper probabilities  and their bounds can be used in three different applications, namely multiple groups inference,  diagnostic accuracy and ranked set sampling.

\subsection{Multiple groups inference} \label{sec.app.1}

The Jonckheere-Terpstra (JT) test is often used to test the null hypothesis that the medians of different populations are equal against the alternative that the values of these medians are in a specific order. In this section, we show how the NPI method presented in this paper can be used in this setting. In the NPI method there is no hypothesis testing; our uncertainty is quantified using lower and upper probabilities. So in this context, we assume that we have $q$ groups, and we have some observed data per group.  NPI provides inference based on the next future observation per group, and then the NPI lower and upper probabilities that these future observations are  ordered in a specific way are obtained as presented in this paper.

To illustrate our method, we use five examples from the literature, these are the same examples used by \citet{TM03} to compare their proposed nonparametric test (NNT) to the Kruskal-Wallis (KW) test, the Jonckheere-Terpstra (JT) test, and the modified version of the Jonckheere-Terpstra (MJT) \citep{Neuh98}. Examples 1-4  correspond to the data given in Table 5 (Example 2) of \citet[p.907]{Neuh98}, Table 6.16 of \citet[p.209]{Daniel78}, Table 6.7 of \citet[p.211]{HW99}, and Table 1 (Replicate 2) of 
\citet[p.589]{SM86}, respectively. Example 5 is the same as Example 4 with 10 added  to each of the observations in the last group. So  Example 1 consists of four groups, Example 2 consists of three groups, Example 3 consists of five groups and Examples 4 and 5 consist of six groups each. The results of these tests are summarised in Table III in \citet{TM03}, the p-values extracted from that table are given in Table \ref{tab.examples1}. The lower and upper bounds (including algorithm A's  approximation) for the  NPI lower and upper  probabilities for the event that $X_{1,n_1+1}<X_{2,n_2+1}<\ldots<X_{q,n_q+1}$, are also given in Table \ref{tab.examples1}. In addition, the corresponding NPI lower and upper  probabilities  calculated from Equations \eqref{eq.per1}-\eqref{eq.per4}, are given in Table \ref{tab.examples2} as a reference.

\begin{figure}
    \centering
    \includegraphics[scale=0.85]{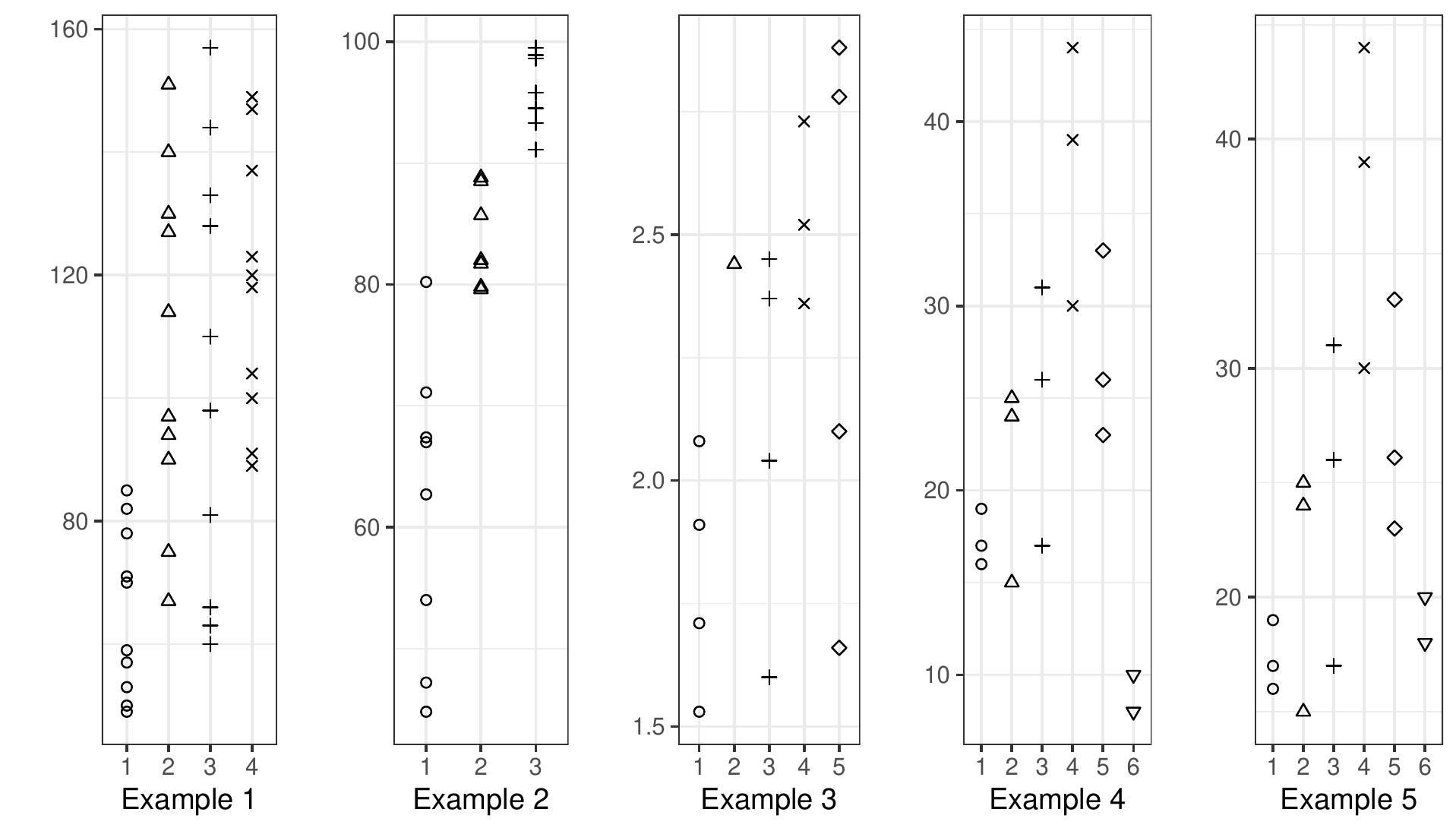}
    \caption{Multiple groups inference examples}
    \label{fig.examples}
\end{figure}

The five data sets  are visualised in Figure \ref{fig.examples} where the different groups are represented on the $x$-axes. From this figure we can see that  Example 2 data set  is reasonably ordered. This is also clear from the significance of all test statistics (KW, JT, MJT and NNT). The NPI lower and upper  probabilities are also large and close to the reference lower and upper  probabilities in Table \ref{tab.examples2}, which indicate a strong evidence that these groups are perfectly ordered. Although from Figure \ref{fig.examples} it seems that the observations from the last three groups in Example 1 are overlapped, e.g.\  observations from group 3 are spread over the range of the other groups, all  test statistics  reject the null hypothesis at  significance level 1\%. The NPI lower and upper probabilities are small and far away from the reference values in  Table \ref{tab.examples2} which suggests no evidence that these groups are perfectly ordered.

For Examples 4 and 5, all test statistics fail to reject the null hypothesis, for the significance level 1\%, with larger p-values for NNT test. The NPI lower probabilities are equal to zero in these two examples and the upper probabilities are close to zero as well suggesting no evidence that these groups are perfectly ordered. For Example 3, only the NNT test is significant at  significance level 1\%. The NPI lower probability is close to zero while the upper probability could indicate a weak evidence that these groups might be perfectly ordered.

\begin{table}[]
    \centering\footnotesize
    \begin{tabular}{c|cccc|ccc|ccc}
        &KW&JT&MJT&NNT& $\underline{P}^L$   & $\underline{P}^A$ & $\underline{P}^U$ & $\overline{P}^L$& $\overline{P}^A$ &  $\overline{P}^U$ \\
       \hline
       Example 1  &0.001&$<0.001$&$<0.001$&0.001& 0.0415 & 0.0578  &  0.0870 & 0.1282& 0.1657 &0.2236  \\
       Example 2 &$<0.001$&$<0.001$&$<0.001$&$<0.001$& 0.5590& 0.5590 & 0.6562& 0.8472& 0.9722 &0.9722 \\
       Example 3 &0.202&0.015&0.021&$<0.001$ & 0 & 0 &0.0160 &0.0300& 0.1140 &0.3270 \\
       Example 4&0.037&0.225&0.331&0.591  & 0&0 &0&0.0033& 0.0189& 0.0820 \\
       Example 5  &0.075&0.066&0.057&0.591& 0& 0& 0& 0.0033& 0.0189& 0.0905 \\
       \hline
    \end{tabular}
    \caption{Multiple groups inference examples}
    \label{tab.examples1}
\end{table}

\begin{table}[]
    \centering\footnotesize
    \begin{tabular}{c|cccc}
      &   $\underline{P}^{L*}$ &   $\underline{P}^{U*}$ & $\overline{P}^{L*}$&  $\overline{P}^{U*}$ \\
       \hline
       Example 1 & 0.5532&0.6830&0.8264 &1  \\
       Example 2 & 0.5833 &0.6806&0.8750&1 \\
       Example 3 & 0 & 0.1920&0.3000&1\\
       Example 4&0.0312&0.1582&0.3164&1 \\
       Example 5  &0.0312&0.1582&0.3164&1 \\
       \hline
    \end{tabular}
    \caption{Multiple groups inference examples}
    \label{tab.examples2}
\end{table}
\subsection{Diagnostic accuracy assessment}
The approach presented in this paper can be used in assessing the accuracy of diagnostic tests.  To this end, we can use the NPI lower and upper probabilities (bounds and approximations) as bounds for the hyper-volume under the  receiver operating characteristic (ROC) hyper-surface, see for example \cite{CMEC2014} for the three groups case where bounds for the volume under the ROC surface are provided. 

\begin{figure}
    \centering
    \includegraphics[scale=0.75]{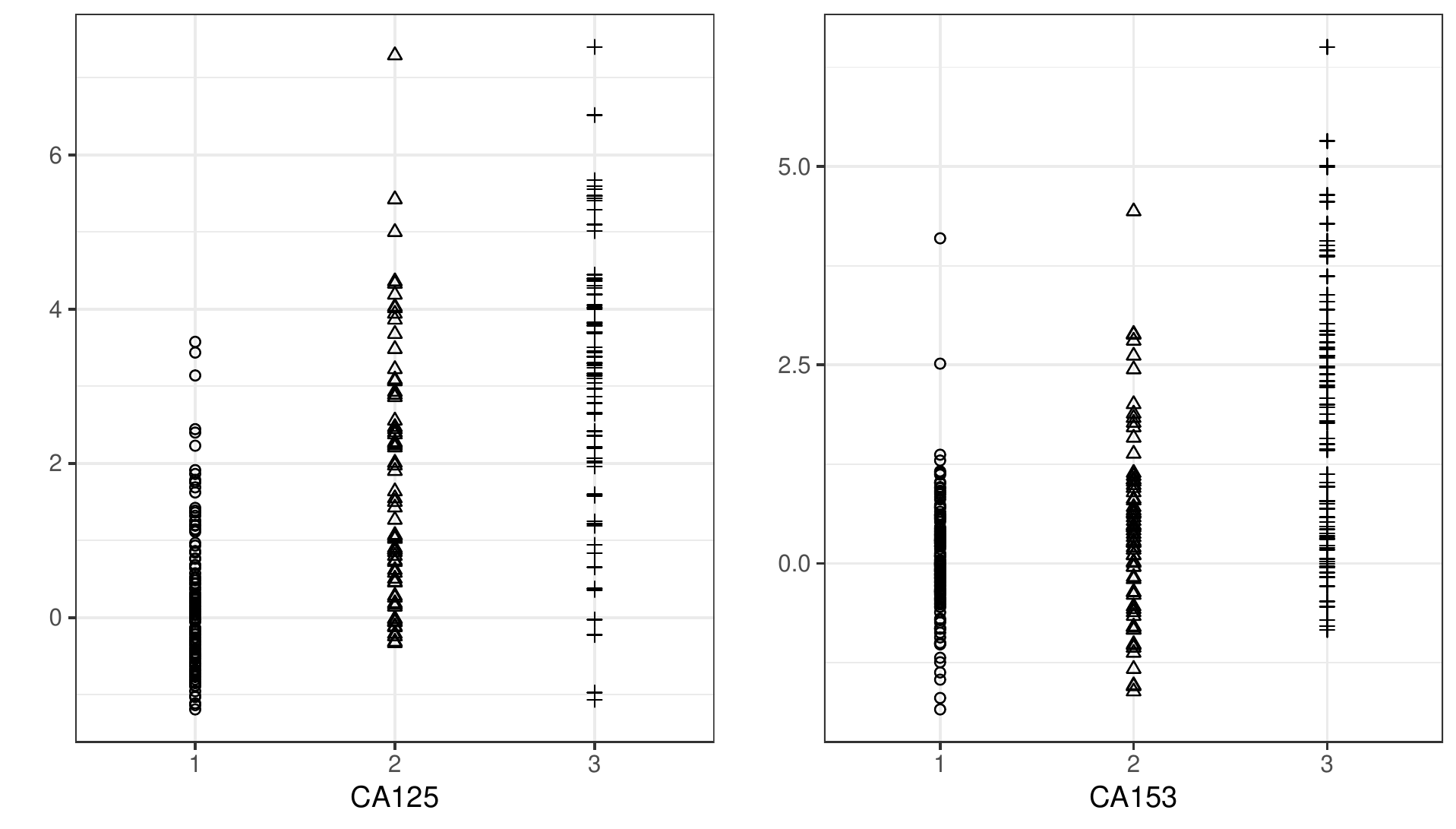}
    \caption{Diagnostic accuracy assessment}
    \label{fig.examples2}
\end{figure}

To illustrate our method, we use a subset of the Pre-PLCO Phase II dataset \citep{bcROC18}, visualised in Figure \ref{fig.examples2}. There are 278 observations in total and three levels of disease status:  benign, early stage and late stage (denoted as 1, 2 and 3, respectively). In this example two biomarkers, CA125 and CA153, are considered in order to compare their diagnostic accuracy. We are interested in the volume under the ROC surface, i.e. $VUS=P(X_1<X_2<X_3)$. To this end, we use the NPI lower and upper probabilities introduced in this paper to infer about the uncertainty of the event of interest that $X_{1,n_1+1}<X_{2,n_2+1}<X_{3,n_3+1}$. So we assume there is one future observation (individual) per group, then we  derive the NPI lower and upper probabilities for the event that the next future individuals from these groups are perfectly ordered. That is  we provide NPI lower and upper bounds for the  volume under the ROC surface. 

The  results  are  summarised  in  Table \ref{tab.examples3}. From this table we can see that the volume under the ROC surface for biomarker CA125 is greater than the the volume under the ROC surface for biomarker CA153, which gives an indication that biomarker CA125 is better than biomarker CA153, as $\underline{P}(\text{CA125}) > \overline{P}(\text{CA153})$. We also notice that the imprecision (the difference between the upper and lower probabilities) is small, this is due to the large sample sizes.

\begin{table}[]
    \centering\footnotesize
    \begin{tabular}{c|ccc|ccc}
      &   $\underline{P}^{L}$ & $\underline{P}$&  $\underline{P}^{U}$ & $\overline{P}^{L}$& $\overline{P}$&  $\overline{P}^{U}$ \\
       \hline
       CA125 & 0.5334&0.5346&0.5466&0.5623&0.5745&0.5759  \\
       CA153 & 0.3315&0.3334&0.3431&0.3559&0.3660&0.3679 \\
       \hline
    \end{tabular}
    \caption{Diagnostic accuracy assessment}
    \label{tab.examples3}
\end{table}

\subsection{Ranked set sampling}
Ranked set sampling,  first proposed by  \cite{McIn52}, is often considered as an alternative to  simple random sampling when the measurement of the characteristic of interest is costly and time consuming \citep{LB08}. Many inferences based on ranked set sampling  assume perfect ranking of the samples. Several nonparametric tests for perfect ranking  have been introduced, see e.g.\  \cite{Frey07}, \cite{LB08} and \cite{VB11}. \cite{CBS04} provided an excellent review of ranked set sampling and its applications. In this section, we show via an example how the method presented in this paper can be used to quantify the uncertainty of perfect ordering. We should distinguish here between the concept of 'perfect ranking' which is the main underlying assumption for many RSS methods and the concept of 'perfect ordering' of a set of future observations as considered in this paper. 

A study has been conducted to investigate the effect of four different sprayer settings on the amount of spray deposit on the leaves of apple trees \citep{Murray00}.  The data set, given in  Table \ref{ex.tab2}, is based on a five-cycle ranked set samples with each cycle being of size $n=5$. This data set has been used by \cite{LB08} in order to compare different nonparametric tests for perfect ranking in ranked set sampling.

\begin{table}[htp]
\caption{Ranked set sample of spray deposit (percentage of cover)}
\begin{center}
\begin{tabular}{|c|ccccc|}
\hline
Ranks  & \multicolumn{5}{c|}{Cycle}\\ 
\cline{2-6}
& 1 & 2 & 3& 4 & 5\\
\hline
1	&	0.3	&	3.9	&	3.4	&	5.1	&	3.2\\
2	&	2.8	&	11.9	&	11.8	&	10.4	&	14.1\\
3	&	24.4	&	12.6	&	13.0	&	19.3	&	13\\
4	&	5.7	&	10.5	&	21.8	&	21	&	25\\
5	&	14.3	&	56.5	&	29.6	&	15	&	22.9\\
\hline
\end{tabular}
\end{center}
\label{ex.tab2}
\end{table}%

For our method, we  group all observations that have rank $j$ as group $X_j$, and we assume that these groups are independent and the observations within these groups are exchangeable. Suppose now we  have a future cycle, based on the information we have from the previous five cycles.  We  are  interested in the lower and upper probabilities for the event that  $X_{1,6}<X_{2,6}<X_{3,6}<X_{4,6}<X_{5,6}$. We also need to assume that $X_{j,n+1}$ is exchangeable with the observations of group $j$, where $j=1,2,3,4,5$. Note that here we  apply NPI differently, in which this future observation is assumed to come from the same process  as the past observations that have rank $j$ as they all have been selected from different cycles. So in this case we apply the $A_{(n_j)}$ assumption per group $X_j$. Applying the method introduced in this paper,  we can calculate the lower and upper bounds for the NPI lower and upper probabilities as $\underline{P}^L=\frac{165}{7776}=0.0212$, $\underline{P}^A=\frac{345}{7776}=0.0444$, $\underline{P}^U=\frac{594}{7776}=0.0764$, 
$\overline{P}^L=\frac{1024}{7776}=0.1317$, $\overline{P}^A=\frac{1384}{7776}=0.1780$ and $\overline{P}^U=\frac{2674}{7776}=0.3439$. 
These values indicate that it is unlikely for the next future observations to be in this specific order, i.e. 
$X_{1,6}<X_{2,6}<X_{3,6}<X_{4,6}<X_{5,6}$.


\section{Concluding remarks}\label{sec.con}
In this paper, we introduced NPI lower and upper probabilities for the event that future observations from multiple groups are  ordered in a specific way. Several applications of the proposed methods are considered including multiple groups inference,  diagnostic accuracy and ranked set sampling. We have also introduced two algorithms to obtain approximations for the exact NPI lower and upper probabilities. 
Algorithm B is preferable as long as it is computationally feasible. At the same time, Algorithm A may still provide a reasonable approximation for the exact NPI lower and upper probabilities in cases with more than three groups, in which Algorithm B may become computationally expensive.

The work presented in this paper can be extended in many ways. For example, when dealing with lifetime data one may need to take censoring into account. NPI for right censored data has been introduced by  \cite{CY04}, further NPI-based inferences for right censored data  have been considered by \cite{Maturi10} including multiple comparisons, precedence testings and competing risks.  Extending the method proposed here to deal with right censored data will open further applications in survival analysis and reliability.   It is of great interest to extend the proposed method to include covariate information. Developing  NPI  for regression-type models is currently in progress, once fully developed, we intend to apply it to such scenarios.  Another way in which this work can be extended is to consider  umbrella alternatives as orderings; for example, consider the event that $X_{1,n_1+1}<\ldots<X_{i,n_i+1}>\ldots> X_{q,n_q+1}$. In this case one needs to be careful on how to assign the probability masses in order to minimise for the lower probability and to maximise for the upper probability. These topics are left for future investigation.

\section*{Acknowledgements}

The author would like to thank Prof.\ Balakrishnan for the stimulating discussions during his visit to Durham in November 2018, and acknowledges receipt of a Durham University Global Engagement Travel Grant supporting this visit. The author would like to thank the two anonymous reviewers whose suggestions helped improve  this paper.



\end{document}